\documentstyle[aps,preprint,prb]{revtex}

\begin{document}

\title{Effects of doping on spin correlations \\ in the
periodic Anderson model} 
\author{J.\ Bon\v ca} 
\address{
Department of Physics,
FMF, University of Ljubljana and J. Stefan Institute,
Ljubljana, Slovenia}
\author {J.\ E.\ Gubernatis} 
\address{
Theoretical Division, 
Los Alamos National Laboratory, Los Alamos, NM 87545} 

\date{ \today}

\maketitle
\begin{abstract}
We studied the effects of hole doping on spin correlations in the
periodic Anderson model, mainly at the full and three-quarters-full
lower bands cases. In the full lower band case, strong anti-ferromagnetic
correlations develop when the on-site repulsive interaction strength
$U$ becomes comparable to the quasi-particle band width. In the
three-quarters full case, a novel kind of spin correlation develops
that is consistent with the resonance between a $(\pi,0)$ and a $(0,\pi)$
spin-density wave. In this state the spins on different sublattices
appear uncorrelated. Hole doping away from the completely full case
rapidly destroys the long-range anti-ferromagnetic correlations, in a
manner reminiscent of the destruction of anti-ferromagnetism in
the Hubbard model. In contrast to the Hubbard model, the doping does
not shift the peak in the magnetic structure factor from the
$(\pi,\pi)$ position.  At dopings intermediate to the full and
three-quarters full cases, only weak spin correlations exist.
\end{abstract}


\section{Introduction}

We report the results of a quantum Monte Carlo study on the effects of
hole doping on spin correlations in the two-dimensional asymmetric
periodic Anderson model. Our undoped state is the insulating state
which exists when the lower band is full. When the on-site Coulomb
repulsion is sufficiently large, this state is also
anti-ferromagnetic. On the one hand, we will argue that the effects of
hole doping on this state are similar to those found in doping studies
of the Hubbard model's half-filled state as we find that the doping
rapidly destroys the anti-ferromagnetism. On the other hand, we will
argue that continued doping to a $3/4$-filling of the lower band
produces a novel ground-state, not reported for the Hubbard model,
that we interpret as a state resonating between two spin-density wave
states. In this state, spins on the different sub-lattices of the
bipartite square lattice are uncorrelated. 

Fewer analytic studies exist for the periodic Anderson model (PAM)
than for the Hubbard model. Yet the published literature is still
vast. In one, two, three, and infinite dimensions for strong and weak
coupling limits of the interaction strengths, many different
investigators have studied this model by using conserving
approximations, the density-matrix renormalization group method, the
Hartree-Fock, slave boson, and dynamical mean-field approximations,
the exact diagonalization method, and various variational methods. If
studied, the effect of doping was found to induce paramagnetic,
ferromagnetic, anti-ferromagnetic, and spiral spin states at various
temperatures and interaction strengths.  Interest in the periodic
Anderson model has traditionally been driven by interests in mixed
valent, heavy fermion, and Kondo insulating materials. Possible
application to these classes of materials motivated the present study.

Far fewer quantum Monte Carlo (QMC) studies exist for the Anderson
model than for the Hubbard model. Most have been for the
one-dimensional symmetric PAM. In fact we are aware of only 2
one-dimensional studies of the asymmetric PAM\cite{saso} and 2
two-dimensional simulations\cite{zhang,vekic} of the symmetric
PAM. Both two-dimensional studies were simulations at low but finite
temperature and focused on the symmetric PAM at
half-filling to avoid the Fermion sign problem which produces
exponentially growing variances in measured properties as the
temperature is lowered, the lattice size is increased, and
particle-hole symmetry is destroyed by doping.

Zhang and Callaway,\cite{zhang} in the earlier paper of the two,
concluded that the properties of the half-filled symmetric PAM had
some qualitative similarities to those of the single-impurity Anderson
model but at low temperatures some additional physics developed. In
the single-impurity model, the electrons on the impurity site become
correlated with the conduction electrons as the temperature is lowered
to the Kondo temperature $T_K$. At temperatures much lower than $T_K$,
the local moment on the impurity is spin compensated by the conduction
electrons, leading to a non-magnetic state.  Zhang and Callaway
observed similar behavior in the lattice model. Short range
correlations between the conduction electrons and the electrons
localized on the f-orbitals screened these moments, and at very low
temperatures these local moments became correlated through their
interaction with the conduction electrons. The build-up of these
correlations was seen in the behavior of the magnetic structure factor
and the uniform magnetic susceptibilities. It is the development of
the anti-ferromagnetic correlations among the f-orbitals that is the
new physics in the periodic Anderson lattice model.\cite{lee}

More recently, Veki\'c et al.\cite{vekic} found for the same model the
same behavior observed by Zhang and Callaway, plus more as they
explicitly sought to examine the competition between the tendency to
anti-ferromagnetic order and to spin disorder.  As noted by
Doniach,\cite {doniach} this competition is between the
Ruderman-Kittel-Kasuya-Yosida (RKKY) interaction trying to order the
f-moments and the Kondo interaction trying to screen them away.  The
Kondo interaction scales as $T_K \sim W e^{-W/J}$, and the RKKY
interaction, as $J_{RKKY} \sim J^2/W$. ($W$ is some band width, and
$J$ is the strength of the anti-ferromagnetic Kondo exchange
interaction between the conduction band and f-electrons.)  When
$J_{RKKY} > T_K$, an anti-ferromagnetically ordered singlet ground
state is expected.  This will occur as long as $J$ is less than a
critical value $J_c$. When $J> J_c$ a Kondo spin-compensated ground
state is expected. For the PAM, $J\sim V^2/U$, where $V$ is the
strength of the hybridization between the f-electrons and the
conduction band electrons and $U$ is the value of the repulsive
on-site Coulomb interaction. Consequently, when $U$ is large, $J$ is
small, and an anti-ferro-magnetically ordered state is expected. As
the value of $U$ is lowered, a cross-over to a disordered ground-state
eventually occurs. The general conclusions of Veki\'c et al. are even
broader: For small values $J$, the
ground state of the PAM is an insulator with long-range
anti-ferromagnetic order characterized by a finite charge gap and
gapless spin excitations. As $J$ increases, the long-range order
is destroyed, and the system exhibits spin-liquid behavior, which is a
disordered spin state with both a spin and charge gap.  When $J1<$
is sufficiently large, the system crosses over to a band-insulating
state, and the size of the spin and charge gaps approach each other.

In this report we used a new ground-state QMC method, the
constrained-path Monte Carlo method (CPMC), to study the properties of
the asymmetric two-dimensional PAM.  This method\cite{shiwei1}
eliminates the Fermion sign problem that plagues most QMC methods for
simulating systems of interacting electrons. It accomplishes this by
imposing an approximate condition constraining the random walk of the
ground state wavefunction. Testing and comparing with
other simulations has demonstrated\cite{shiwei1,shiwei2,mariana,janez}
that the estimated energy and the predicted correlation functions are
very accurate.  Consequently, with this method we can more
effectively study doped systems than we could with other QMC
methods.\cite{eugene}

We limited our numerical study of the PAM to select ranges of interaction
strengths and dopings. In the undoped state, we positioned the energy
of the f-orbital into the lower band and increased the strength of the
on-site Coulomb interaction until on the average one electron occupied
each f-orbital with a nearly saturated magnetic moment. From the work
of Veki\'c et al., we expect a gap in the density of states and
long-range anti-ferromagnetic order among the moments on the
f-orbitals.  Hole doping changes the nature of the indirect exchange
interaction and thus the nature of the spin correlations.  For the PAM, doping
also rapidly destroys the anti-ferromagnetic state, and sufficient
doping produces a novel state which we call the resonating
spin-density-wave (RSDW) state. Our main purpose is to illustrate
these results.

In the next Section we describe the PAM, emphasizing its features when
the on-site interaction strength $U$ is zero. These features suggest
the possibility of a magnetic instability of the non-interacting
electron gas at several electron densities. Then in Section~III, we
give a qualitative description of the CPMC method. Details have been
reported elsewhere.\cite{shiwei1} In Section~IV, we present our
numerical results. We emphasize the wavenumber dependence of the
electron density and the spin-spin correlation function. Lastly, in
Section~V, we present summarizing remarks and suggestions for future
work.

\section{Periodic Anderson Model}

For the Hamiltonian describing the PAM we took
\begin{eqnarray}
  H &=& -t\sum_{\langle i,j \rangle,\sigma} 
       (d_{i,\sigma}^\dagger d_{j,\sigma}+d_{j,\sigma}^\dagger d_{i,\sigma})
        +V\sum_{i,\sigma} 
       (d_{i,\sigma}^\dagger f_{i,\sigma}+f_{i,\sigma}^\dagger d_{i,\sigma})
       \nonumber \\
    & & \quad\quad +\epsilon_f\sum_{i,\sigma}n_{i,\sigma}^f
        +\frac{1}{2} U \sum_{i,\sigma}n_{i,\sigma}^fn_{i,-\sigma}^f
\label{eq:pam}
\end{eqnarray}
where the creation and destruction operators create and destroy
d-electrons on sites of a square lattice and f-electrons on orbitals
associated with these sites.  $n_{i,\sigma}^f=f_{i,\sigma}^\dagger
f_{i,\sigma}$ is the number operator for f-electrons.  Elsewhere we
will use a similar notation to denote quantities like
$n_{i,\sigma}^d=d_{i,\sigma}^\dagger d_{i,\sigma}$, the number
operator for d electrons. The lattice has $N$ sites, and hopping only
occurs between between neighboring lattice sites and between a lattice
site and its orbital. We used periodic boundary conditions and took
$t=1$.

From (\ref{eq:pam}) we define $H_{00}$ and $H_{0}$, the resulting
Hamiltonians when $V=U=0$ and $U=0$. $H_{00}$ has two energy bands,
each holding up to $N$ electrons of each spin $\sigma$. One band is
dispersionless with a value of $\epsilon_f$. The other is dispersive
with $2N$ states labeled by the wavevector ${\bf k}=(k_x,k_y)$ and
spin $\sigma$ and is given by
\begin{equation}
  e_\sigma({\bf k}) = -2\bigl[\cos(k_x)+\cos(k_y)\bigr]
\end{equation}
Thus for $N$ lattice sites there are $4N$ available energy states. A
half-filled system corresponds to $2N$ electrons.

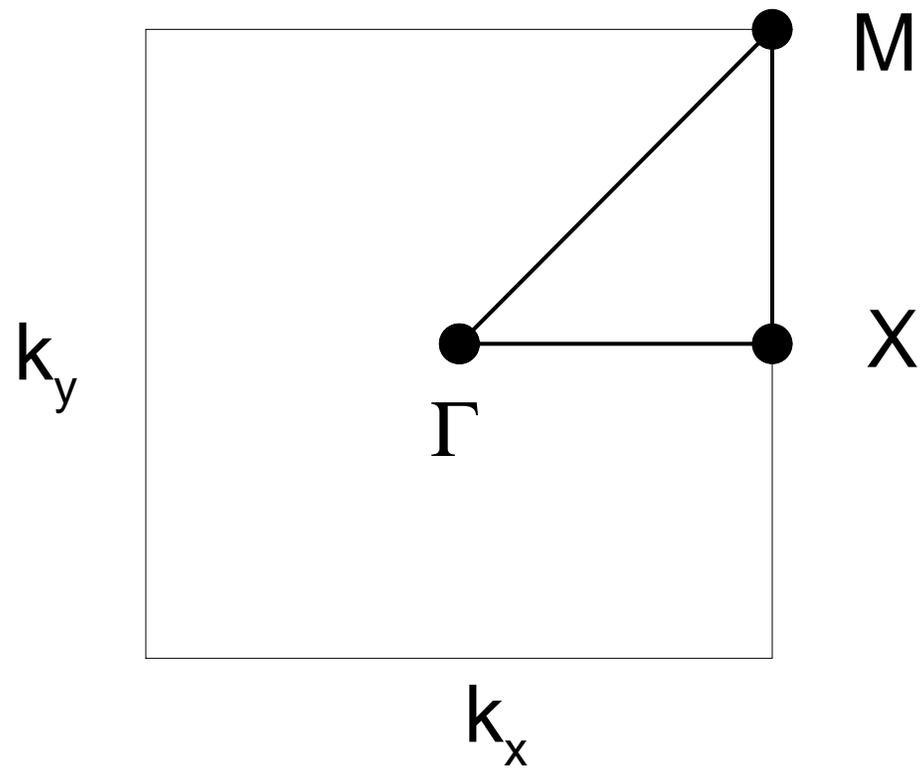
\begin{figure}
\caption{The Brillouin zone of a square lattice. The points $\Gamma$, X, and M
are high symmetry points. The irreducible part of the zone is the
triangular region they define.
\label{fig1}}
\end{figure}

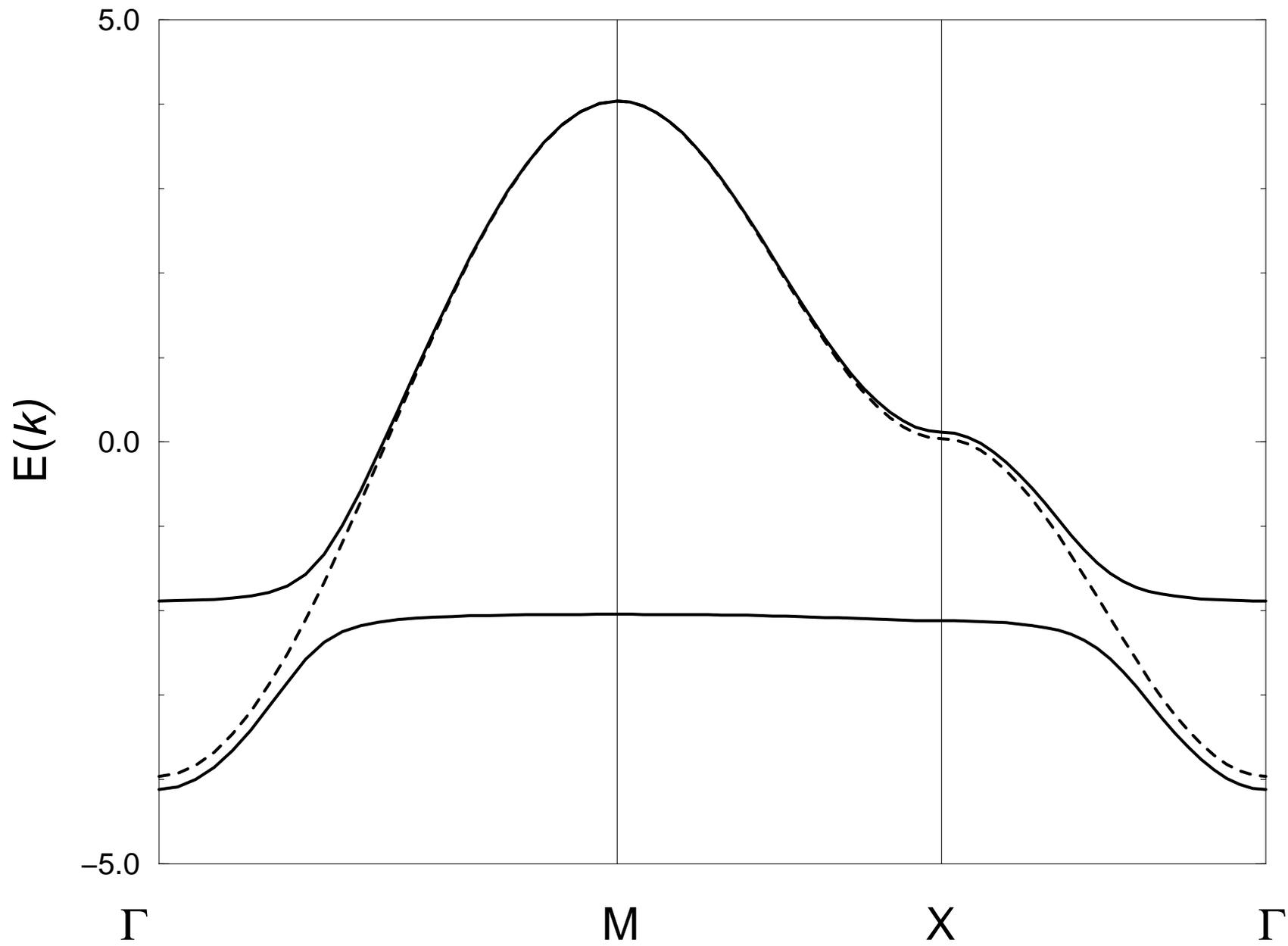
\begin{figure}
\caption{The solid lines are the band structure of the non-interacting
model for $\epsilon_f=-2$ and $V=\frac{1}{2}$. The dashed line is the
tight-binding band shifted to match the value of the upper band at the
point M.
\label{fig2}} 
\end{figure}

$H_0$ has two dispersive bands
\begin{equation}
 E_\sigma^\pm({\bf k})=\frac{1}{2} \Biggl[
   e_\sigma({\bf k})+\epsilon_f \pm \sqrt{(e_\sigma({\bf k})
                                            -\epsilon_f)^2+4V^2}
   \Biggr]
\end{equation}
separated by a gap
\begin{equation}
 \Delta = E_\sigma^{+}(0,0)-E_\sigma^{-}(\pi,\pi)
        = -4 + \frac{1}{2}\Biggl[
        \sqrt{(4+\epsilon_f)^2+4V^2} + \sqrt{(4-\epsilon_f)^2+4V^2}
        \Biggr]
\end{equation}
The operators which create quasi-particles in the lower and
upper bands are of the form
\begin{eqnarray}
 \alpha_{{\bf k},\sigma}^\dagger 
    &=& \sum_i \biggl(Y_{{\bf k}i} f_{i,\sigma}^\dagger
                     +X_{{\bf k}i} d_{i,\sigma}^\dagger\biggr)
 \nonumber\\
 \beta_{{\bf k},\sigma}^\dagger 
    &=& \sum_i \biggl(X_{{\bf k}i} f_{i,\sigma}^\dagger
                     -Y_{{\bf k}i}  d_{i,\sigma}^\dagger\biggr)
\label{alphak}
\end{eqnarray}
and depending on the relative magnitudes of the matrices $X$ and $Y$, a
band can be f-like or d-like. If one band is f-like, then the other
must be d-like. At half-filling, only the states in the lower band are
filled. When this band itself is only half or more filled, it shows little
dispersion among the states near the Fermi surface ${\bf k}_F$,
meaning there is a large density of states at ${\bf k}_F$ and the
nearby band states are f-like. By contrast, the upper-band resembles
the tight-binding band $e_\sigma({\bf k})$ and is thus d-like. The
Brillouin zone for the square lattice is shown in Fig.~\ref{fig1}. A
plot of both bands and the tight-binding band along directions between
high symmetry points in this zone is shown in Fig.~\ref{fig2}.  The
tight-binding band was shifted so it and $E_{+}({\bf k})$ are equal at
$(\pi,\pi)$.

All our simulations were done for hole dopings to less than
half-filling. In the non-interacting case half-filling corresponds to
a full lower band. {\it Because the way we dope, we found it more
convenient, from this point on, to characterized our results in terms of
the fractional filling of the lower band\/}. Also because of the way we
dope, the properties of the lower band are obviously particularly important.
The lower (valance) band has a width of
\begin{equation}
 W = E_\sigma^{-}(\pi,\pi)-E_\sigma^{-}(0,0)
        = 4 - \frac{1}{2}\Biggl[
        \sqrt{(4+\epsilon_f)^2+4V^2} - \sqrt{(4-\epsilon_f)^2+4V^2}
        \Biggr]
\end{equation}        
We only considered $\epsilon_f=-2$ and $V=0.5$, and for these values,
$\Delta=0.16$ and $W = 2.08$. Thus we have a narrow band, with an
enhanced density of states near the Fermi surface, into which we will
place the electrons and then induce strong correlations by adding
strong repulsive Coulomb interactions. We also considered only a
$6\times 6$ lattice size. In terms of computational effort this size
is roughly equivalent to simulating a $12\times 12$ Hubbard model.

Often the symmetric version of the PAM is studied. In this case,
$\epsilon_f = -\frac{1}{2}U$ and at half-filling particle-hole
symmetry exits. As $U$ is varied so is the band-structure of the
non-interacting problem. We chose to keep that structure fixed and
vary the interaction strength. Over the range of parameters used, the
dominance of $T_K$ is replaced by $J_{RKKY}$ as $U$ is increased from
0. Accordingly, the ground-state is expected to go from a spin disordered
to anti-ferromagnetically ordered one.\cite{vekic}

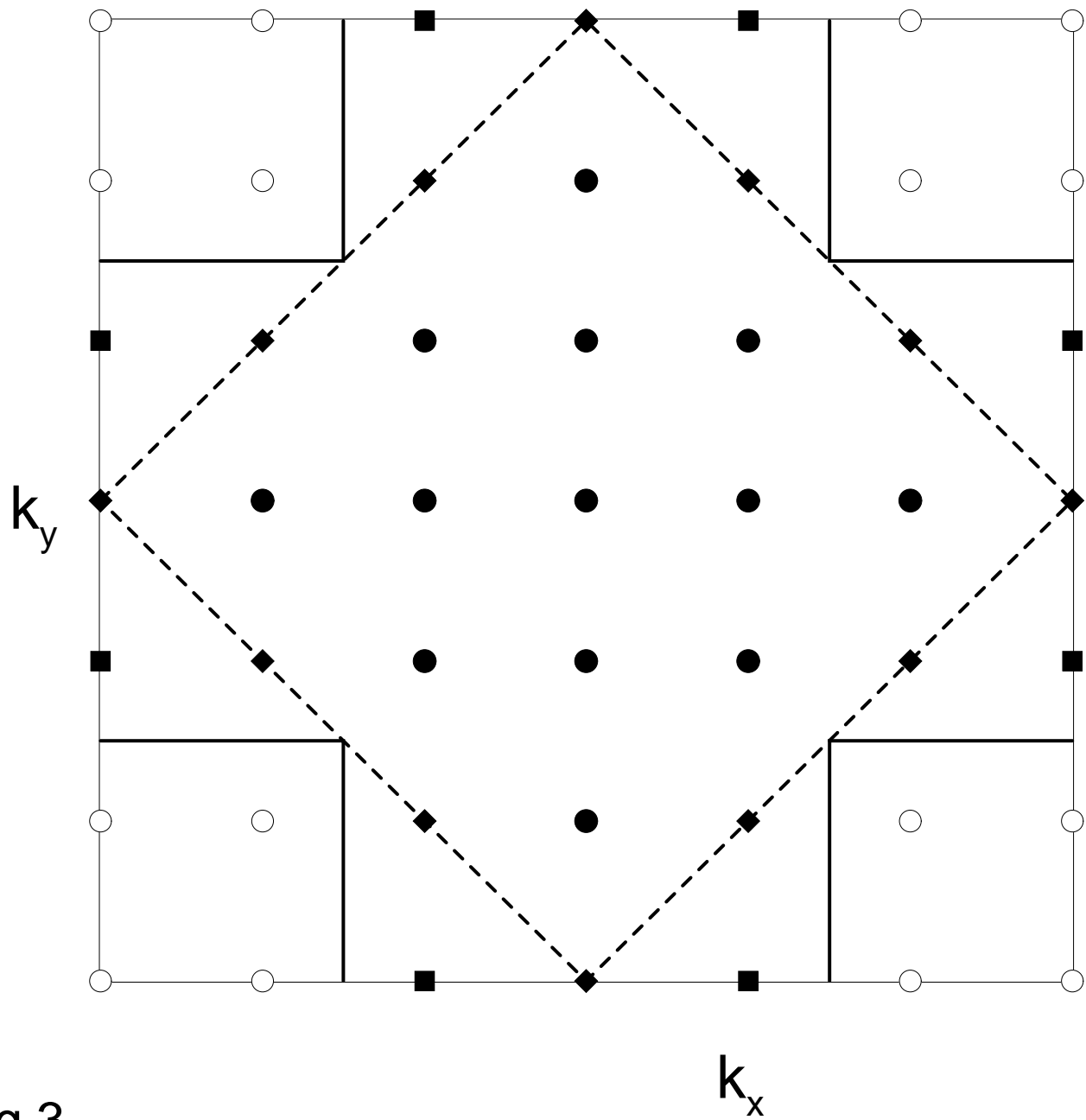
\begin{figure} 
\caption{The points in the first Brillouin zone for a $6\times 6$
lattice. The dashed lines is the fully nested Fermi surface for the
half-filled case. The thick solid lines represent lines of nesting for
a ${\bf Q}=(\pi,0)$ spin-density wave. Three-quarters filling
corresponds to double occupancy of the points denoted by filled
markers.
\label{fig3}} 
\end{figure}

The ground state of $H_0$ does not show long-ranged spatial spin
correlations, but at certain electron fillings the Fermi surface may
be unstable towards the development of a spin-density wave when
$U\not=0$. Such instabilities arise if the nesting conditions,
$E_\sigma^{-}({\bf k})= E_\sigma^{-}({\bf k'}+{\bf Q}) =
E_\sigma^{-}({\bf k}_F) \equiv E_F$ and $E_\sigma^{-}({\bf k'})=E_F$,
are satisfied.\cite{enz} For a commensurate state, ${\bf Q}$ is a high
symmetry point on the boundary of the Brillouin zone and equals
one-half of a reciprocal lattice vector. For a square lattice there
are two such points, $(0,\pi)$ and $(\pi,\pi)$, and the nesting
conditions lead to the set of thick lines in the Brillouin zone shown
in Fig.~\ref{fig3}. If portions of the Fermi surface approximate these
lines, an instability of the Fermi surface is possible. (For the lower
band, the critical fillings are $\frac{1}{4}$, $\frac{1}{2}$, and
$\frac{3}{4}$. In Fig.~\ref{fig3}, we only show the nesting
conditions relative to $\frac{1}{2}$, and $\frac{3}{4}$ fillings.) In
the $6\times 6$ non-interacting problem, three-quarters
filling is a closed-shell case with double occupancy of all points
designated by a solid marker. Half-filling is an open shell case with
double occupancy of the solid circles and the remaining occupancy
being some linear combination of half the solid diamonds.  From
Fig.~\ref{fig2}, one sees that the Fermi energy of the
$\frac{3}{4}$ and completely full states are nearly equal.

When the lower band is $\frac{1}{4}$-filled, the number of electronic
quasi-particles is much less than the number of lattice sites, making it
unlikely that the Coulomb interaction $U$ will induce an instability,
because of the unlikely double occupancy of the f-levels.  Also
the density of states will most likely remain free-electron-like and
small. We did not investigate this filling. At $\frac{1}{2}$-filling,
the number of quasi-particles equals the number of lattice sites and
perfect nesting,\cite{enz} reminiscent of the Hubbard model, occurs and
a similar anti-ferromagnetic insulating state is expected. We also did
not investigate this filling at this time. We focused on fillings of
$\frac{3}{4}$ and higher.

At $\frac{3}{4}$-filling new possibilities exist. The likelihood of double
occupancy is large because the number of quasi-particles is much larger
than the number of lattice sites, and the density of states at $E_F$ is
large because of the dispersionless character of the band. If the
interaction is strong, each f-state on the average will be occupied by
a single electron with a nearly fully developed magnetic moment, and
these moments will interact anti-ferromagnetically.  The questions are,
``Does a magnetic instability characterized by ${\bf Q}=(\pi,0)$
develop?'' and ``What is the nature of this state?''

When the lower band is full, a band gap exists.  For a sufficiently
large value of $U$, each f-orbital on the average will be singly
occupied, the spins of these electrons will be arranged
anti-ferromagnetically, and the gap will be enlarged.  In this paper
we will study the evolution of this state as we hole dope it towards
the $\frac{3}{4}$-filled case where a magnetic state of a different
character develops.

\section{Constrained-Path Monte Carlo Method}

Our numerical method is extensively described and benchmarked
elsewhere.\cite{shiwei1} Here we only discuss its basic approximation.
In the CPMC method, the ground-state wave function $|\Psi_0\rangle$ is
projected from a known initial wave function $|\Psi_T\rangle$ by a
branching random walk in an over-complete space of Slater determinants
$|\phi\rangle$.  In such a space, we can write $|\Psi_0\rangle =
\sum_\phi \chi(\phi) |\phi\rangle$.  The random walk produces an
ensemble of $|\phi\rangle$, called random walkers, which represent
$|\Psi_0\rangle$ in the sense that their distribution is a Monte Carlo
sampling of $\chi(\phi)$, that is, a sampling of the ground-state wave
function.

To completely specify the ground-state wave function, only
determinants satisfying $\langle\Psi_0|\phi\rangle>0$ are needed
because $|\Psi_0\rangle$ resides in either of two degenerate halves of
the Slater determinant space, separated by a nodal plane ${\cal N}$
that is defined by $\langle\Psi_0|\phi\rangle=0$.  The sign problem
occurs because walkers can cross ${\cal N}$ as their orbitals evolve
continuously in the random walk. Asymptotically they populate the two
halves equally.  If ${\cal N}$ were known, we would simply constrain
the random walk to one half of the space and obtain an exact solution
of Schr\"odinger's equation.  In the constrained-path QMC method,
without {\it a priori\/} knowledge of ${\cal N}$, we use a trial wave
function $|\Psi_T\rangle$ and require $\langle\Psi_T|\phi\rangle>0$.
The random walk again solves Schr\"odinger's equation in determinant
space, but under an approximate boundary-condition.  This is what is
called the constrained-path approximation.

The quality of the calculation clearly depends on the quality of the
trial wave function $|\Psi_T\rangle$. Since the constraint only
involves the overall sign of its overlap with any determinant
$|\phi\rangle$, it seems reasonable to expect the results to show some
insensitivity to $|\Psi_T\rangle$.  Through extensive benchmarking on
the Hubbard model, it has been found that simple choices of this
function can give very good
results.\cite{shiwei1,shiwei2,mariana,janez}

Besides as starting point and as a condition constraining a random
walker, we also use $|\Psi_T\rangle$ as an importance
function. Specifically we use $\langle\Psi_T|\phi\rangle$ to bias the
random walk into those parts of Slater determinant space that have a
large overlap with the trial state. For all three uses of
$|\Psi_T\rangle$, it clearly is advantageous to have $|\Psi_T\rangle$
approximate $|\Psi_0\rangle$ as closely as possible. Only in the
constraining of the path does $|\Psi_T\rangle \not= |\Psi_0\rangle$
in general generate an approximation.

All the calculations reported here are done with periodic boundary
conditions.  Mostly, we study closed shell cases, for which the
corresponding free-electron wave function is non-degenerate and
translationally invariant. In these cases, the free-electron wave
function, represented by a single Slater determinant, is used as the
trial wave function $|\psi_T\rangle$.  (The use of an unrestricted
Hartree-Fock wave function as $|\psi_T\rangle$ produced no significant
improvement in the results).

In particular, we represented the trial wavefunction as a single
Slater determinant whose columns are the $N_\sigma$ single-particle
orbitals obtained from the exact solution of $H_0$.  We chose the
orbitals with lowest energies given by
$E_{-}^\sigma({\bf k})$ and filled them up to a desired number of
electrons $N_e$.
\begin{equation}
 \vert \psi_T \rangle = \prod_{\bf k,\sigma} \alpha_{{\bf
 k},\sigma}^\dagger \vert 0 \rangle,
\end{equation}
where $\vert 0 \rangle$ represents a vacuum for electrons.  Since our
calculations were performed at or below a full lower band, only states
from the lower band were used to construct the trial
wavefunction.

In a typical run we set the average number of random walkers to 400.
We performed 2000 Monte Carlo sweeps before we taking measurements,
and we made the measurements in 40 blocks of 400 steps. By choosing
$\Delta \tau = 0.05$, we reduced the systematic error associated with
the Trotter approximation to be smaller than the statistical error. In
measuring correlation functions, we performed between 20 to 40
back-propagation steps. The number of up and down electrons was always
chosen equal $N_\uparrow=N_\downarrow=N_e/2$.

\section{Results}

Our simulations were done for a $6 \times 6$ lattice and lower-band
fillings of $1$, $\frac{3}{4}$ and several values in between. At
these fillings, we varied the repulsive on-site Coulomb interaction
$U$ to witness how the properties of a narrow band of non-interacting
quasi-particles change when the interaction becomes large and how
doping affects these properties. In particular we examined the effects
of the interaction and doping on the electron and spin densities and
the correlations between these densities.

\subsection{Filled Lower Band}

We describe the electronic distribution in several ways. One is by its
local (site) values $\langle n_i^f \rangle$ and $\langle n_i^d
\rangle$. Another is by its momentum distribution
\begin{equation}
  \langle n({\bf k})\rangle  = \langle n^\alpha({\bf k})\rangle 
      + \langle n^\beta({\bf k})\rangle  
      =\sum_\sigma\Bigl[ \langle n_{{\bf k},\sigma}^\alpha\rangle  
      + \langle n_{{\bf k},\sigma}^\beta\rangle \Bigr]
\end{equation}
An still another is by its total occupancies of the lower  and upper bands
\begin{eqnarray}
 \langle n^\alpha \rangle
       &=& \frac{1}{N}\sum_{{\bf k}}\langle n^\alpha({\bf k})\rangle\nonumber\\
 \langle n^\beta \rangle 
       &=& \frac{1}{N}\sum_{{\bf k}}\langle n^\beta({\bf k})\rangle 
\end{eqnarray}
Comparing and contrasting these different quantities as a function of
the interaction strength is often very informative.

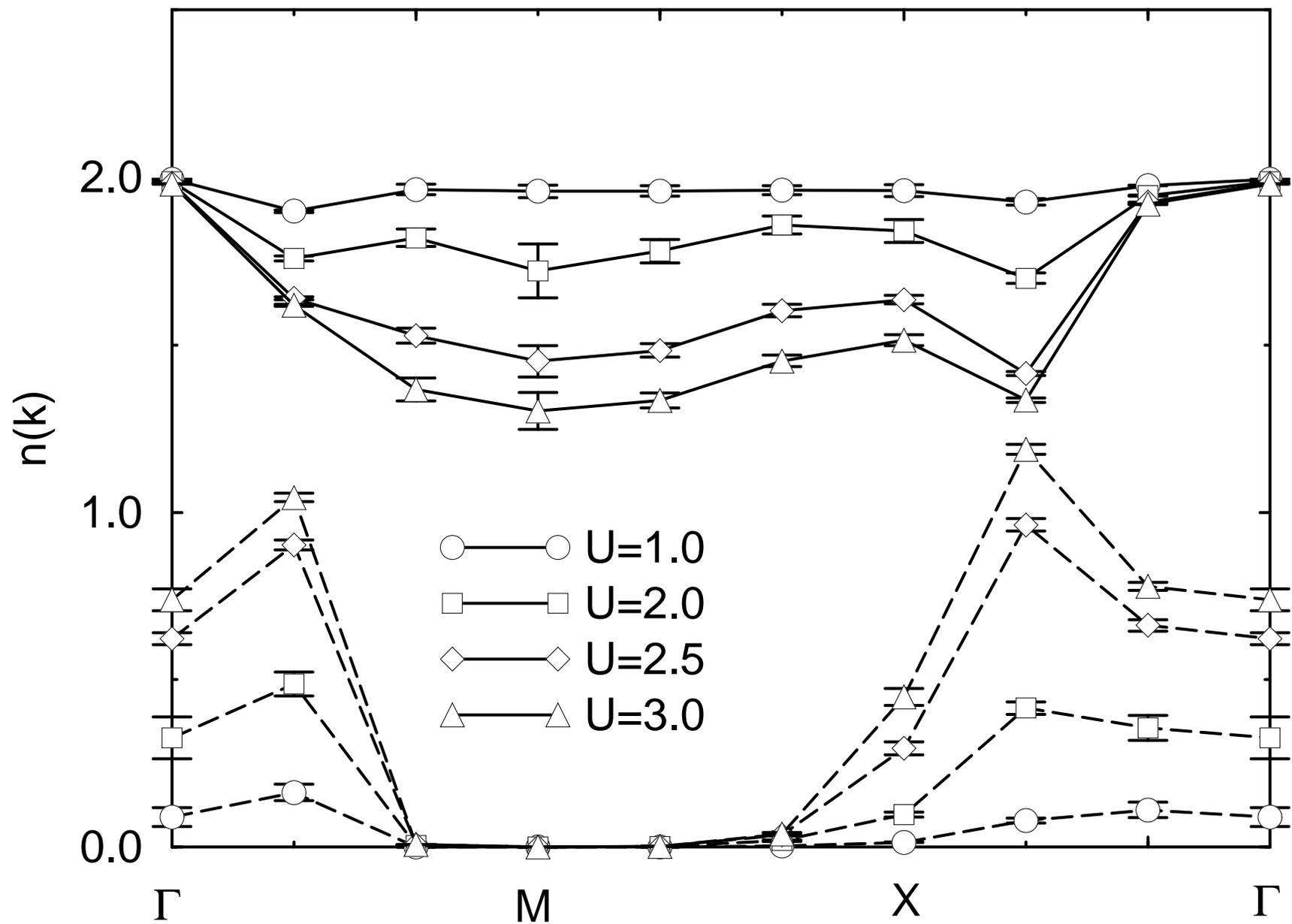
\begin{figure} 
\caption{Electron momentum density as a function of ${\bf k}$ for the
filled lower band as a function of $U$. The solid lines are for
$n_\alpha({\bf k})$ and the dashed lines for $n_\beta({\bf k})$.
\label{fig4}}
\end{figure}

When $U=0$ and the lower band is completely full, $ \langle
n^\alpha({\bf k})\rangle =2$ and $\langle n^\beta({\bf k})\rangle
=0$. When $U>0$, this uniformity changes, but we find only small
changes. Near the $\Gamma$ point in the Brillouin zone, $\langle
n({\bf k})\rangle $ becomes greater than 2.  When this occurs,
$n^\beta({\bf k})$ has become greater than zero at values of ${\bf k}$
where $\langle n^\alpha({\bf k})\rangle $ is nearly 2.  In
Fig.~\ref{fig4}, we plot $\langle n^\alpha({\bf k})\rangle $ and
$\langle n^\beta({\bf k})\rangle $ along the lines connecting the high
symmetry points in the irreducible part of the Brillouin zone.  By
plotting the pieces of $\langle n({\bf k})\rangle $ separately, we get
an enhanced picture of how the quasi-particle picture changes.  What
Fig.~\ref{fig4} suggests, for example, is turning on the repulsive
interaction moves some of the short wavelength electron momentum to
longer wavelengths in presumably higher energy states. In other words,
as $U$ is increased, some of the itinerant character of the
non-interacting quasi-particles is changed to a more localized one and
examining the site-dependent expectation values of the electron
occupancy becomes informative.

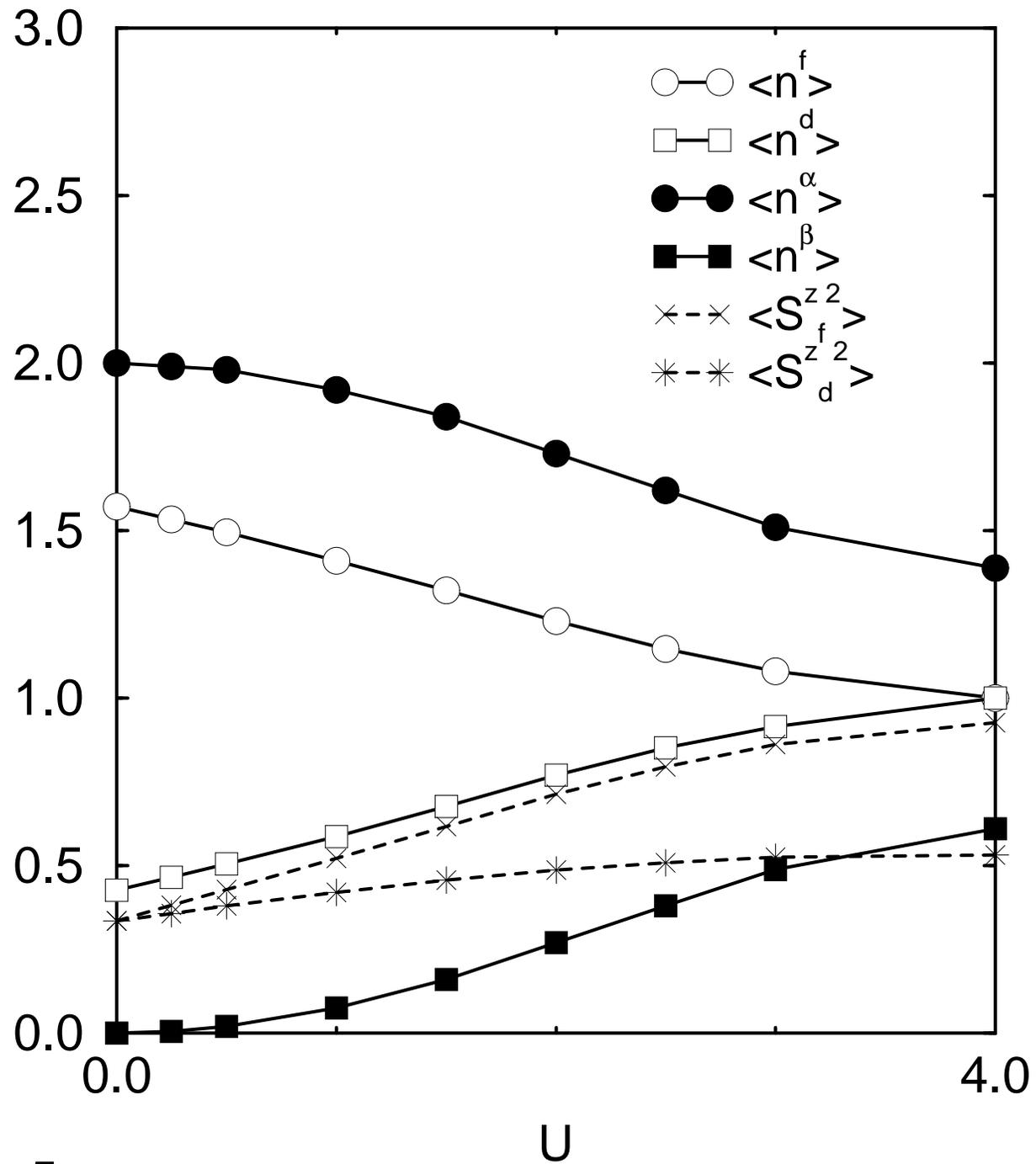
\begin{figure} 
\caption{For the filled lower band, $\langle n_i^d \rangle$, $\langle
n_i^f \rangle$, $\langle n^\alpha \rangle$, $\langle n^\beta \rangle$,
$\langle S_f^z(i)^2\rangle$, and $\langle S_d^z(i)^2\rangle$ as a
function of $U$.
\label{fig5}} 
\end{figure}

This transition is more explicitly seen in Fig.~\ref{fig5} which shows a
nearly saturated moment
\begin{equation}
     \langle S_f^z(i)^2 \rangle = 
     \langle (n_{i,\uparrow}^f-n_{i,\downarrow}^f)^2 \rangle \approx 1
\end{equation}
developing on each f-orbital as $U$ is increased.  This observation
with the additional observation of $\langle n_i^f\rangle \approx 1$ and
the use of the algebraic identity
\begin{equation}
 \langle S_f^z(i)^{2} \rangle = \langle n_i^f \rangle  
         - 2 \langle n_{i,\uparrow}^f n_{i,\downarrow}^f \rangle
\end{equation}
implies that $\langle n_{i,\uparrow}^f n_{i,\downarrow}^f \rangle =0$
and thus demonstrates that a large $U$ localizes a single electron on
each f-orbital. The picture for the d-electrons differs. Here, when
$\langle n_i^d \rangle \approx 1$, $\langle
S_d^{z}(i)^{2}\rangle\approx 0.5$, and the identity
\begin{equation}
 \langle S_d^z(i)^{2} \rangle = \langle n_i^d \rangle - 2 \langle
         n_{i,\uparrow}^d n_{i,\downarrow}^d \rangle
\end{equation}
implies that $\langle n_{i,\uparrow}^d n_{i,\downarrow}^d
\rangle\approx \frac{1}{4}$. Thus, the d-electrons do not sit one per
site but instead show some itinerant character.

The spin-spin correlation function
\begin{equation}
  S_{ff}({\bf k}) = \frac{1}{N}\sum_{i,j} 
  e^{i {\bf k}\cdot({\bf R}_i-{\bf R}_j)}
  \langle (n_{i,\uparrow}^f-n_{i\downarrow}^f)
       (n_{j,\uparrow}^f-n_{j,\downarrow}^f)\rangle
  =\frac{1}{N}\sum_{i,j}
  e^{i {\bf k}\cdot({\bf R}_i-{\bf R}_j)}
  \langle S_f^z(i)S_f^z(j)\rangle
\end{equation}
plotted in Fig.~\ref{fig6}a, shows a strong enhancement of
$S_{ff}(\pi,\pi)$ occurring as $U$ is increased. In contrast, the
enhancement to $S_{dd}(\pi,\pi)$, shown in Fig.~\ref{fig6}b, is  much
smaller.  Additional insight follows from the spatial correlation
functions. The function $\langle S_f^z(i)S_f^z(j)\rangle$ for $i \not=
j$ shows long-range anti-ferromagnetic correlations. The magnitude
of these correlations are a factor of 20 to 30 larger than those shown
by $\langle S_d^z(i)S_d^z(j)\rangle$ (for $i \not= j$). The
d-electrons are behaving as a collection of weakly anti-ferromagnet,
rather itinerant electrons.

A large $U$ separates the f and d-electrons into two coupled
anti-ferromagnetic layers. To study the correlations between these
layers, we calculated $\langle S_f^z(i)S_d^z(j)\rangle$.
Figure~\ref{fig7} shows this function for a small and a large value of
$U$. For the small value of $U$, when $i=j$, the correlation between
the hybridized sites is negative; otherwise, it decays rapidly with
the distance away from the orbital site in a manner reminiscent of the
decaying correlations found in the single impurity Anderson
model.\cite{gubernatis} For a large value of $U$, when $i=j$, the
correlation between hybridized sites is again negative; otherwise, it
shows long-range anti-ferromagnetic oscillations. In general, the
magnitude of spatial spin correlations between the d and f-electrons is
about an order of magnitude larger than those between the d-electrons.

For all values of $U$ studied, the expectation value of the net total
magnetization was zero: 
\begin{equation}
  \langle S^z \rangle = \sum_i \langle S_d^z(i)+S_f^z(i) \rangle=0
\end{equation}
From this and the fact that the ground state is an eigenstate of the
total magnetization, the following constraints on the spatial spin
correlations functions hold and were observed in the results of the
simulations
\begin{equation}
       \sum_i \langle S_f^z(i)S_f^z(j)\rangle 
    =  \sum_i \langle S_d^z(i)S_d^z(j)\rangle 
    = -\sum_i \langle S_f^z(i)S_d^z(j)\rangle
\label{eq:singlet}
\end{equation}
Rewriting these equations 
\begin{eqnarray}
     \langle S_f^z(j)^2\rangle 
     &= &
    -\sum_i \langle S_f^z(i)S_d^z(j)\rangle
    -\sum_{i\not=j} \langle S_f^z(i)S_f^z(j)\rangle\nonumber\\
     \langle S_d^z(j)^2\rangle 
     &= &
    -\sum_i \langle S_f^z(i)S_d^z(j) \rangle
    -\sum_{i\not=j} \langle S_d^z(i)S_d^z(j)\rangle
\label{eq:screening}
\end{eqnarray}
shows that both the d and f-electrons participate in the spin
compensation of the local d and f-moments and that this compensation is
qualitatively different in the PAM from the spin compensation in the
single-impurity model.\cite{gubernatis,blankenbecler} In the
single-impurity model, the impurity spin compensation formula
is\cite{gubernatis}
\begin{equation}
     \langle S_f^z(j)^2\rangle = 
    -\sum_i \langle S_d^z(i)S_f^z(j)\rangle
\label{eq:kondoscreening}
\end{equation}
This equation is not even approximately obeyed for the PAM.  In the
PAM compensation of the local moment definitely involves the moments
on the other orbitals, not just the d electrons.  This possibility was
suggested by Nozi\'eres\cite{nozieres} for non-dilute Kondo alloys and
noted by several authors\cite{jarrell,kyung} in the context of
dynamical mean-field calculations for the PAM. We remark that spin
compensation holds for all dopings and values of $U$ as long as
$\langle S^z \rangle =0$, and while it is consistent with a singlet
ground state, it is not proof of a singlet ground
state.\cite{blankenbecler} In the Kondo regime, as in the
single-impurity model, short-range spin correlations act to compensate
the f-moments; in the anti-ferromagnetic regime, long-range
correlations act.

Similar correlations functions can be computed for the charge. For
example,
\begin{equation}
  C_{ff}({\bf k}) = \frac{1}{N}\sum_{i,j} 
  e^{i {\bf k}\cdot({\bf R}_i-{\bf R}_j)}\biggl[
  \langle (n_{i,\uparrow}^f+n_{i,\downarrow}^f)
       (n_{j,\uparrow}^f+n_{j,\downarrow}^f)\rangle -
  \langle n_{i,\uparrow}^f+n_{i,\downarrow}^f\rangle
  \langle n_{j,\uparrow}^f+n_{j,\downarrow}^f\rangle\biggr]
\end{equation}
The magnitude of these correlation functions is 2 to 3 orders smaller
than the magnitude of $S_{ff}(\pi,\pi)$. Shown in Figs.~\ref{fig8}a and
\ref{fig8}b are the charge-charge correlation functions for the filled
lower band case. For the d-electrons, increasing $U$ increases the
magnitude of the charge correlations slightly, and these correlations
are the strongest near $(\pi,\pi)$. Increasing $U$ suppresses the
f-electron charge correlations. In general, the changes in these
correlations are small relative to the $U=0$ values. The behavior of
the charge-charge correlation functions as a function of $U$ is
qualitatively and quantitatively the same at other fillings and so
these functions will not be discussed further.

\subsection{Incommensurate Fillings}

When we start hole doping away from the filled lower band, the
principal effect is the eventual destruction of the anti-ferromagnetic
correlations. For small $U$, where these correlations do not exist,
doping does not qualitatively change the behavior of the physical
quantities we computed. The principal change is a reduction in the
value of $\langle n^\beta({\bf k}) \rangle$, that is, the removal of
charge from the upper band.  The lattice is a ``charge reservoir'' for
the orbitals. When $U$ is large, doping reduces the value of $\langle
n_i^d\rangle$, that is, removes electrons from the lattice sites.

At large $U$ the most significant change caused by the doping is the
immediate and significant reduction of the magnitude of the long-range
spatial correlations of the spins. To be more specific, fixing $U$ at
a value for which we found anti-ferromagnetism
in the full lower band case, then to our statistical accuracy we found that
this anti-ferromagnetism disappeared by a filling of 33 up and 33
down electrons. Strong local moments on the orbitals, however,
remained as in general we find that for a fixed value of $U$ the
magnitude of the moment is only a weak function of the doping. Doping
mainly affects the nature of the indirect exchange mechanism that
induces the anti-ferromagnetism. In contrast to the two-dimensional
Hubbard model, the doping did not shift the peak in the spin-spin
correlation from the $(\pi,\pi)$ point to the incommensurate points
$(\pi\pm q,0)$ and $(0,\pi\pm q)$.\cite{imada}

After the anti-ferromagnetic correlations are destroyed, continued
doping changes the qualitative features of the computed quantities
change very little. At large $U$, local moments still exist, the
orbitals are singly-occupied, and the remaining electrons show
free-electron-like spin-spin and charge-charge correlations.
Short-range spin and charge correlations exist as evidenced by the
respective correlation functions showing broad peaks at values of
${\bf k}$ not equal to $(0,0)$ or $(\pi,0)$.  The height of these
peaks are orders of magnitude smaller than the height of the
$(\pi,\pi)$ peak peak that existed for the large values of $U$ in the
full lower band case.  In Figs.~\ref{fig9} and \ref{fig10} we
present some of the properties of a system with 31 up and 31 down
electrons.

\subsection{Three-quarters Filled}

At large $U$ and $\frac{3}{4}$-filling, a peak in the spin-spin
correlation function dramatically appears at ${\bf k}=(\pi,0)$. We
will attribute this peak to the formation of a resonating pair of
${\bf Q}=(\pi,0)$ and ${\bf Q}'=(0,\pi)$ spin-density waves. The
behavior of the system exhibits several other characteristics
different from those found at other fillings. At this filling, when
$U=0$, $\langle n_i^f \rangle
\approx 1$ and $\langle n_i^d \rangle \approx 0.5$, and double occupancy
of the f-orbitals and lattice sites is small. A large $U$ is not
needed to induce single occupancy of the f-orbitals. Increasing $U$
actually slightly decreases the f occupancy and slightly increases the d
occupancy. 

In Fig.~\ref{fig11}, we show the electron momentum distribution. As
$U$ is increased, two effects occur. The first effect is the expected
development of a nearly saturated moment on each f-orbital
(Fig.~\ref{fig12}). The second effect is the apparent disappearance of
the quasi-particle residue\cite{pines}
\begin{equation}
  Z({\bf k_F}) = n({\bf k}\rightarrow {\bf k}_F^+)
               - n({\bf k}\rightarrow {\bf k}_F^-)
\end{equation}
The Fermi surface is becoming less apparent, if it still exists. 

The ${\bf Q}=(\pi,0)$ peak in the spin-spin correlation is shown in
Fig.~\ref{fig13}.  The remaining correlation functions are similar in
magnitude and features as for the other dopings. However, when $S_{ff}({\bf
k})$ peaks, there was no peak in $S_{dd}({\bf k})$.  Although we will
not explicitly demonstrate that the peak in $S_{ff}({\bf k})$
signifies a state of long-range order, its presence is consistent with an
ordered state where the spins are aligned ferromagnetically
in rows and anti-ferromagnetically in columns. There are two such
alignments and they correspond to a commensurate spin-density wave
characterized by ${\bf Q}=(\pi,0)$ and its symmetry equivalent
$(0,\pi)$. We will now argue that the state we see is a linear
combination of both spin-density wave states, a state which we call a
resonating spin-density wave (RSDW) state.

If we look at the spatial spin-spin correlation function $\langle
S_f^z(i)S_f^z(j)\rangle$, we find strong correlations existing between
points on the same sub-lattice but very weak ones between points on
the different sub-lattices of the bi-partite square lattice. We argue that
the correlations between sub-lattices are in fact zero. We base the
argument on the additional computation of the transverse spin-spin
correlation function, $\frac{1}{2}\langle S_f^+(i)S_f^-(j) +
S_f^-(i)S_f^+(j)\rangle$. As noted by Hirsch,\cite{hirsch} this
correlation function can often be computed with smaller statistical
error than the longitudinal function. We found this to be the case
here. The predicted correlations (Table~1) between the sub-lattices
were five times smaller than those found with the longitudinal
correlation function and were the same size as the statistical
error. Hence to the accuracy of our simulation they are zero.

The observed state is consistent with the ground-state resonating
between two spin-density wave states, represented schematically for a
$3\times 3$ lattice as
\begin{equation}
\begin{array}{ccc}\uparrow   & \uparrow   & \uparrow   \\
                  \downarrow & \downarrow & \downarrow \\
                  \uparrow   & \uparrow   & \uparrow   \end{array} +
\begin{array}{ccc}\uparrow   & \downarrow & \uparrow   \\   
                  \uparrow   & \downarrow & \uparrow   \\
                  \uparrow   & \downarrow & \uparrow   \end{array} =
\begin{array}{ccc}\uparrow   & 0          & \uparrow   \\   
                  0          & \downarrow & 0          \\
                  \uparrow   & 0          & \uparrow   \end{array}
\label{eq:rsdw1}
\end{equation}
We could equally well depict the process as 
\begin{equation}
\begin{array}{ccc}\uparrow   & \uparrow   & \uparrow   \\
                  \downarrow & \downarrow & \downarrow \\
                  \uparrow   & \uparrow   & \uparrow   \end{array} -
\begin{array}{ccc}\uparrow   & \downarrow & \uparrow   \\   
                  \uparrow   & \downarrow & \uparrow   \\
                  \uparrow   & \downarrow & \uparrow   \end{array} =
\begin{array}{ccc}
                  0          & \uparrow   & 0          \\
                  \downarrow & 0          & \downarrow \\ 
                  0          & \uparrow   & 0          \end{array}
\label{eq:rsdw2}
\end{equation}
Either process produces diagonally-crossed anti-ferromagnetic chains
separated by diagonally-crossed chains with no spins. We argue that
$\langle S_f^z(i)\rangle =0$ is produced by linear combinations of the
two SDW states and their translational symmetry equivalents. We also
observe the equality of the longitudinal and transverse correlation
functions on an element by element basis which suggests that the
f-electrons are in a singlet state.

As we hole dope away from this unusual state, the peak height at
$(\pi,0)$ decreases and the peak width broadens. We did not
extensively investigate these fillings.

\section{Conclusions}
 
We studied the effects of hole doping on spin correlations in the
periodic Anderson model, mainly at the full and three-quarters-full
lower bands cases. In the full lower band case, strong anti-ferromagnetism
develops when the $U$ becomes comparable to the quasi-particle band
width. In the three-quarters filled case, a novel kind of spin
correlation develops between the moments of the f-orbitals that is
consistent with the resonance between a $(\pi,0)$ and a $(0,\pi)$
spin-density wave. In this state, which we call a resonant
spin-density wave (RSDW) state, we also find that the spins on the
different sublattices of the bipartite square lattice appear
uncorrelated. Hole doping away from the completely full case rapidly
destroys the long-range anti-ferromagnetic correlations, in a manner
reminiscent of the destruction of anti-ferromagnetism in the Hubbard
model. In contrast to the Hubbard model, the doping does not shift the
peak in the magnetic structure factor from the $(\pi,\pi)$ position.
Particle and hole doping away from the three-quarters full case was
not extensively studied, but doping appears to destroy this state
relatively rapidly. At dopings intermediate to the full and
three-quarters full case, only weak spin correlations exist.

At a given doping, we found that increasing $U$ develops a strong
local magnetic moment on each f-orbital, mixes the upper and lower
band quasi-particles, promotes the single occupancy of the f-orbitals,
and pushes the remainder of the electrons onto the lattice
sites. These lattice electrons appear free-electron-like. In the
filled band case, they showed only minor correlations with the
anti-ferromagnetic order that develops among the moments on the
orbitals; at three-quarters filling, they show no correlation with the
magnetic structure on the orbitals. When $\langle S^z \rangle =0$, we
noted that both the d and f-electrons participate in the screening
of the d an f-moments.

We did not do the finite size scaling necessary to establish
long-range order at the two central fillings we studied. For the full
lower band case, Veki\'c et al.\cite{vekic} did this study for the
symmetric model and demonstrated long-range order.  We see no reason
why long-range order would be absent when we increased $U$ and moved
the model towards the symmetric case. We are also leaving the
finite-size scaling study of the state at three-quarters filling to
future work, when we can study this state and the effects of doping on
it in more detail. Whether the RSDW state is one of long-range order
is an open question.  We did do several short simulations for a
$8\times 8$ lattice, which requires about an order of magnitude more
computer time than for the $6\times 6$ case. The peak in $S_{ff}({\bf
k})$ at $(\pi,0)$ persisted, but our statistical error increased more
than its height. A Hartree-Fock calculation also exhibited the
peak. We remark that for the $8\times 8$ lattice, $\frac{3}{4}$-filling
is not a closed shell case, and for such cases our measured quantities
in general have larger statistical errors than for closed shell cases.
We also are leaving to a future study the presence of charge and spin
gaps.  Although delicate, the computation of these gaps is within the
capability of our numerical method. In the future, it would also be
interesting to explore the half-full lower band case. This filling
corresponds to one electron per site and seems marginal for the
development of strong electronic correlation phenomena.  If $U$ is
large, some question to ask include, What is the nature of induced
indirect exchange interaction?  Do any anti-ferromagnetic correlations
ever develop? and Does the PAM act like a half-filled Hubbard
model?\cite{nozieres,jarrell}

\section*{Acknowledgments}
This work was supported by the Department of Energy. Very helpful
conversations with B. Brandow, M. Guerrero, M. Jarrell, and G. Ortiz
are gratefully acknowledged.

\begin{figure} 
\caption{Spin-spin correlation functions as a function of
wavenumber and $U$ for the filled band case. (a) $S_{ff}({\bf k})$;
(b) $S_{dd}({\bf k})$.
\label{fig6}} 
\end{figure} 

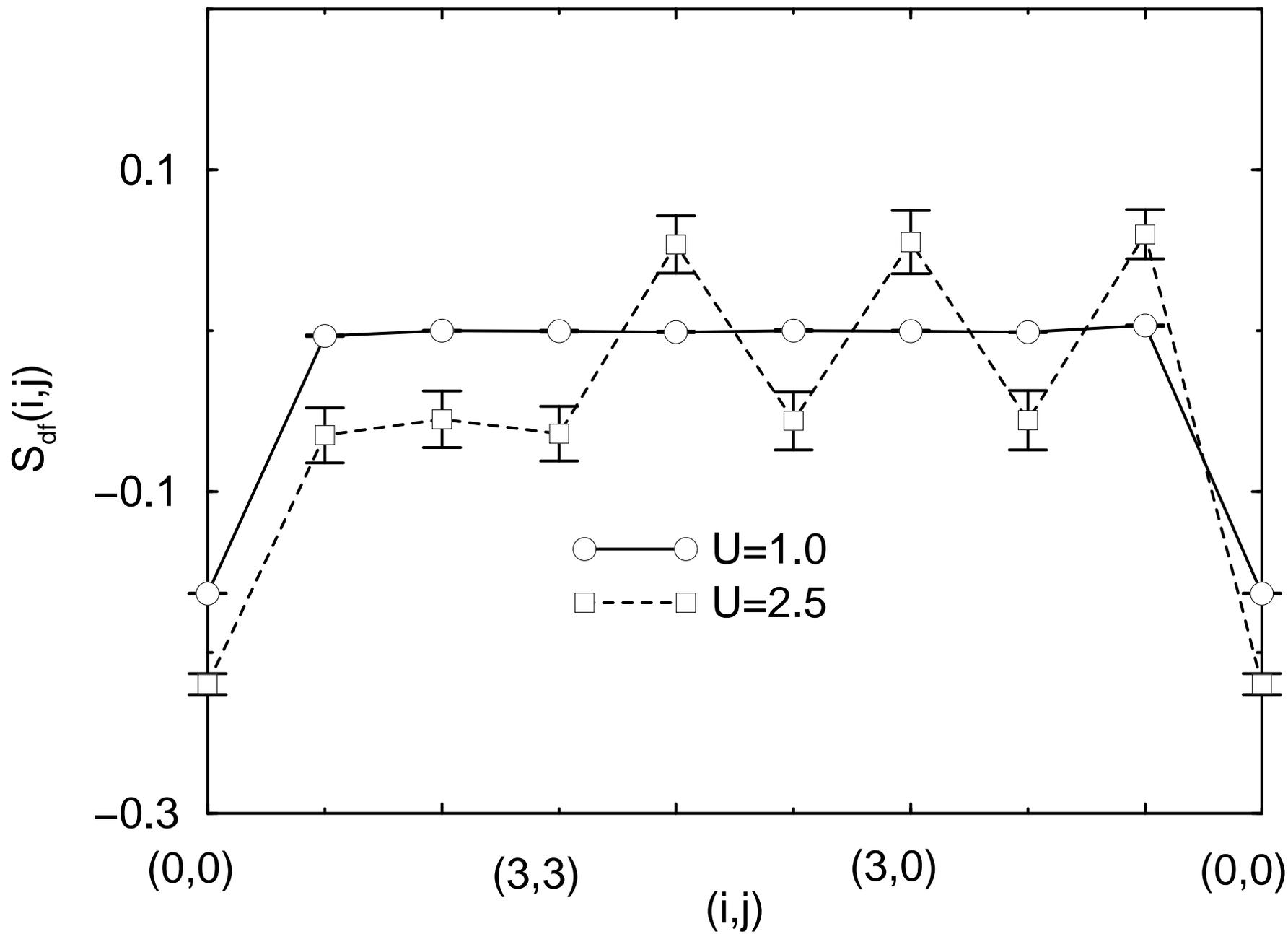
\begin{figure}
\caption{Spin-spin correlation function $\langle
S_d^z(i)S_f^z(j) \rangle$ for the filled band case as a function of
distance between high symmetry points in the unit cell. (a) $U=1$; (b)
$U=2.5$.
\label{fig7}} 
\end{figure} 

\begin{figure} 
\caption{Charge-charge correlation functions as a function of
wavenumber and $U$ for the filled band case. (a) $C_{ff}({\bf k})$;
(b) $C_{dd}({\bf k})$.
\label{fig8}} 
\end{figure} 

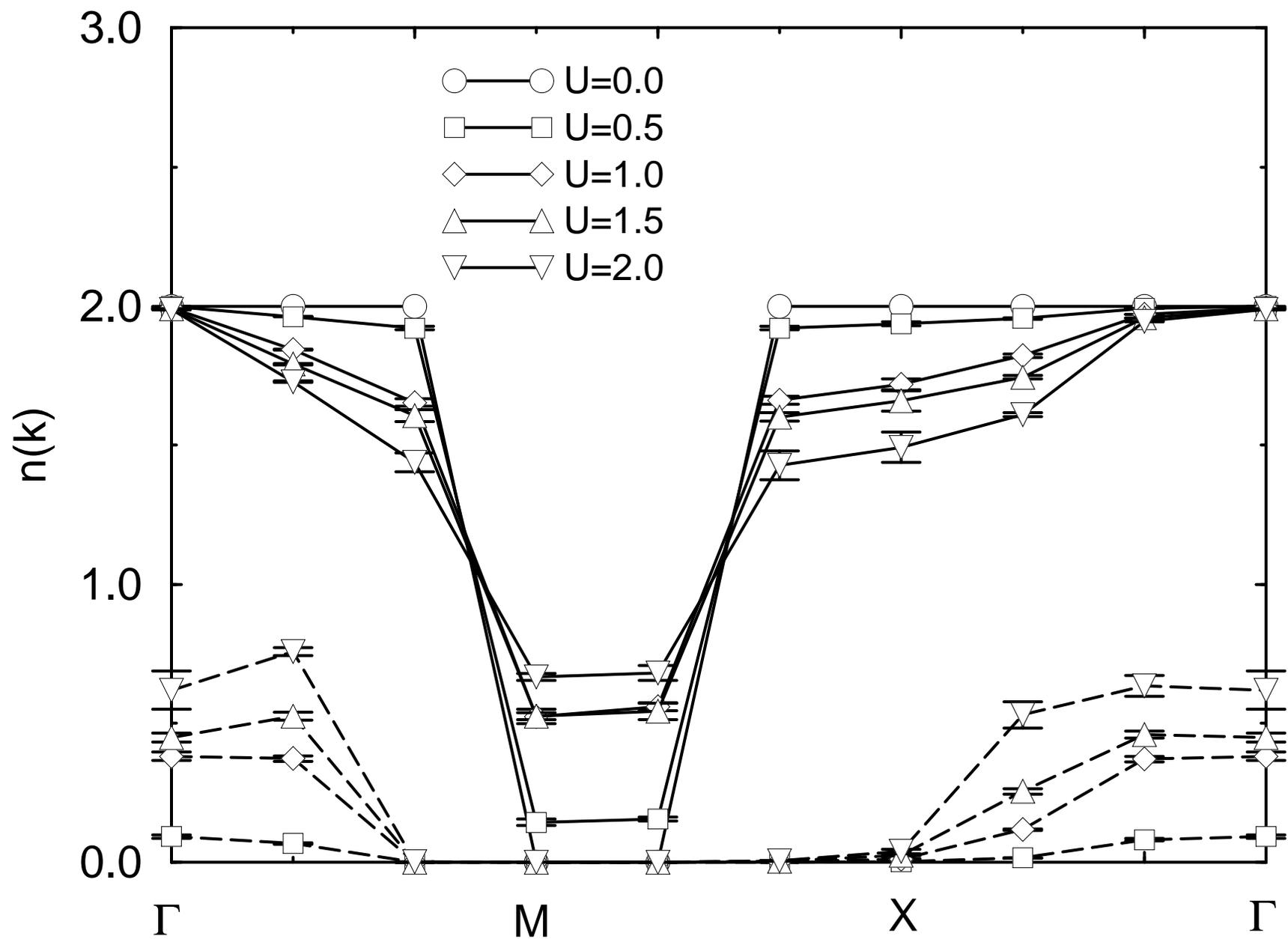
\begin{figure} 
\caption{Electron momentum density as a function of ${\bf k}$ for a
filling of 31 up and 31 down electrons as a function of $U$. The solid
lines are for $n_\alpha({\bf k})$ and the dashed lines for
$n_\beta({\bf k})$.
\label{fig9}} 
\end{figure} 

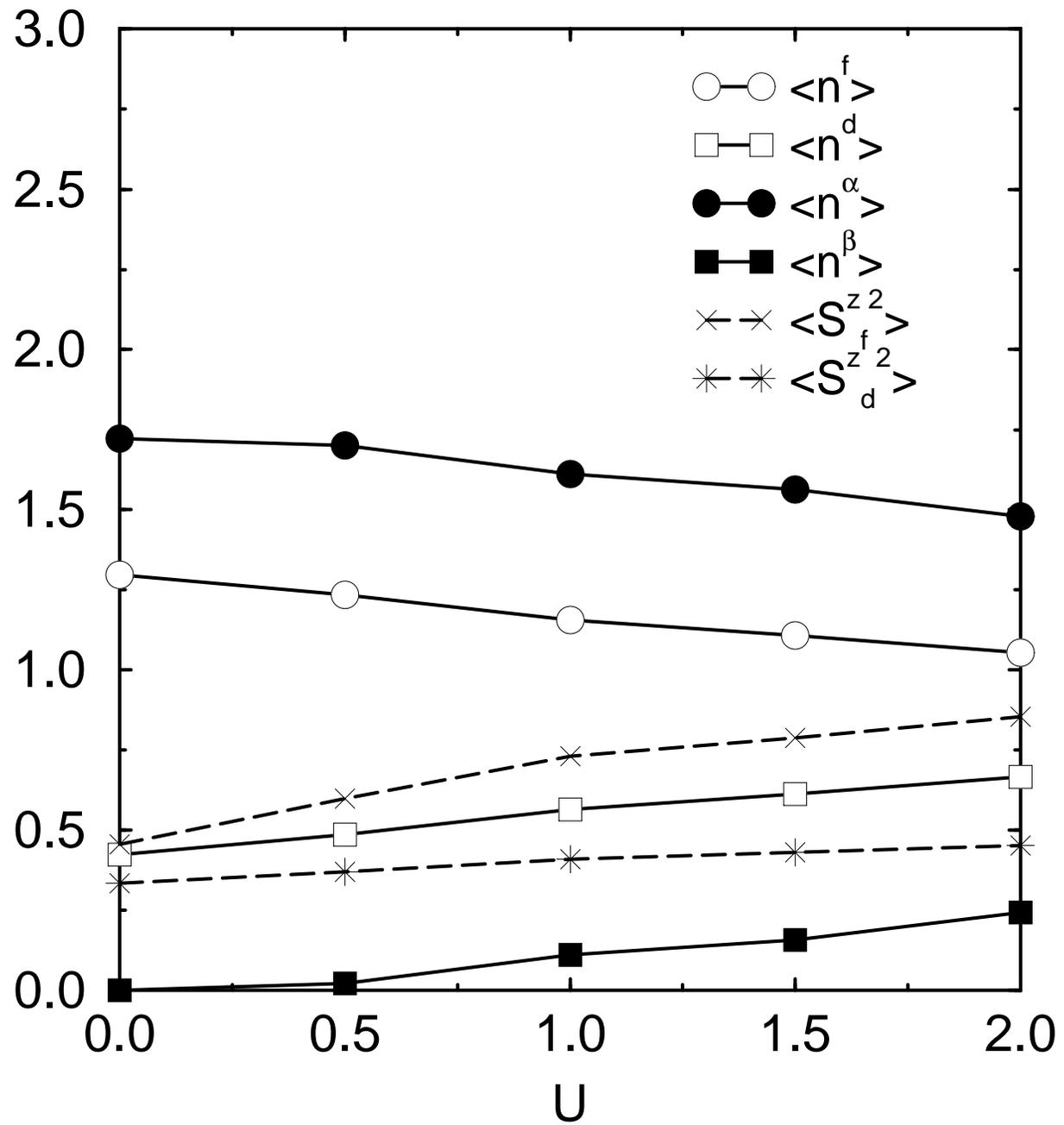
\begin{figure} 
\caption{For 31 up and 31 down electrons, $\langle n_i^d \rangle$, $\langle
n_i^f \rangle$, $\langle n^\alpha \rangle$, $\langle n^\beta \rangle$,
$\langle S_f^z(i)^2\rangle$, and $\langle S_d^z(i)^2\rangle$ as a
function of $U$.
\label{fig10}} 
\end{figure} 

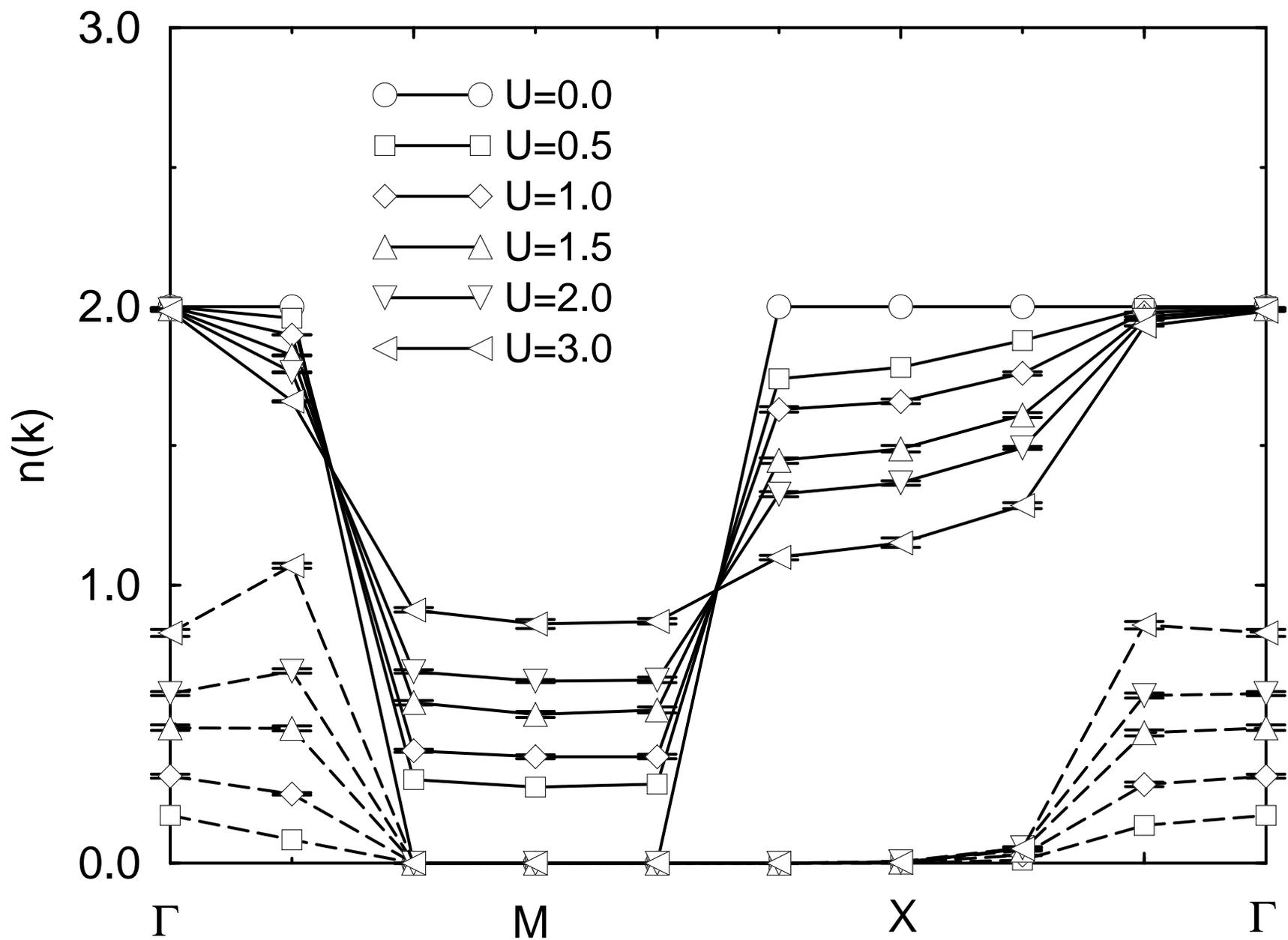
\begin{figure} 
\caption{Electron momentum density as a function of ${\bf k}$ for a
$\frac{3}{4}$ filling as a function of $U$. The solid
lines are for $n_\alpha({\bf k})$ and the dashed lines for
$n_\beta({\bf k})$. 
\label{fig11}} 
\end{figure} 

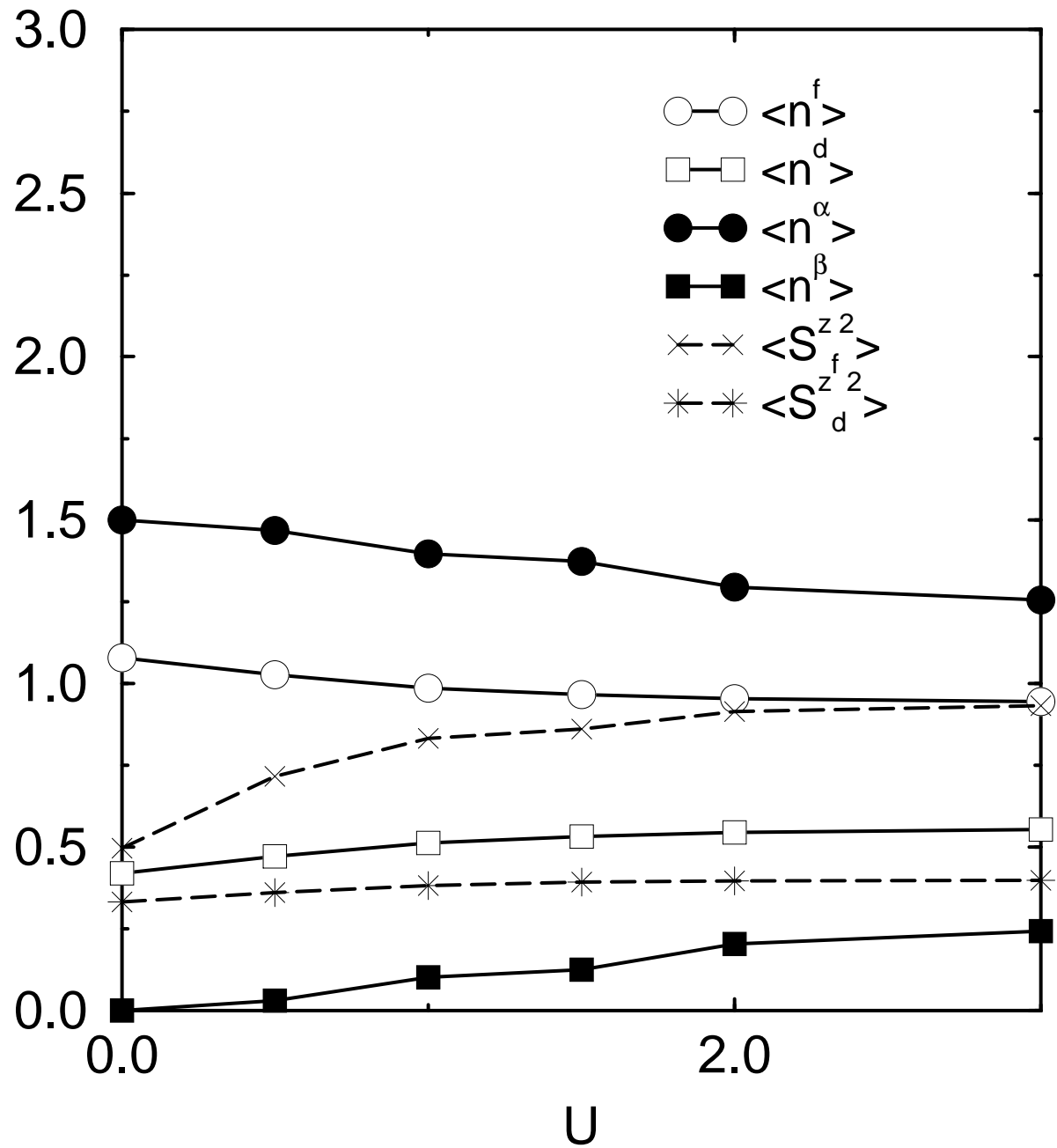
\begin{figure} 
\caption{For the $\frac{3}{4}$-filled lower band, $\langle n_i^d
\rangle$, $\langle n_i^f \rangle$, $\langle n^\alpha \rangle$,
$\langle n^\beta \rangle$, $\langle S_f^z(i)^2\rangle$, 
and $\langle S_d^z(i)^2\rangle$ as a
function of $U$. 
\label{fig12}} 
\end{figure} 

\begin{figure} 
\caption{Spin-spin correlation functions as a function of
wavenumber and $U$ for the $\frac{3}{4}$-filled band case. (a)
$S_{ff}({\bf k})$; (b) $S_{dd}({\bf k})$.
\label{fig13}} 
\end{figure}


\begin{table}
\caption{The  spatial spin-spin correlation function $\langle
S_f^z(i)S_f^z(j)\rangle$ for a $6 \times 6$ lattice. Given are only
the lattices sites in one quadrant of the super cell and these sites
labeled by the point $(i,j)$ starting from the cell center
$(0,0)$. The numbers in parenthesis are the estimated statistical
errors of the last digits.}
\begin{tabular}{ccccc}
~~$(i,j)$~~&    0      &      1     &     2     &     3 \\ \hline
0      & 0.893(0)  & 0.000(6)   & 0.152(38) &-0.006(4)  \\
1      & 0.002(7)  &-0.284(45)  & 0.000(17) &-0.151(50) \\
2      & 0.157(78) & 0.003(11)  & 0.109(30) &-0.001(6)  \\
3      &-0.010(7)  &-0.131(26)  &-0.003(7)  &-0.087(35) \\
\end{tabular}
\label{sisj}
\end{table}

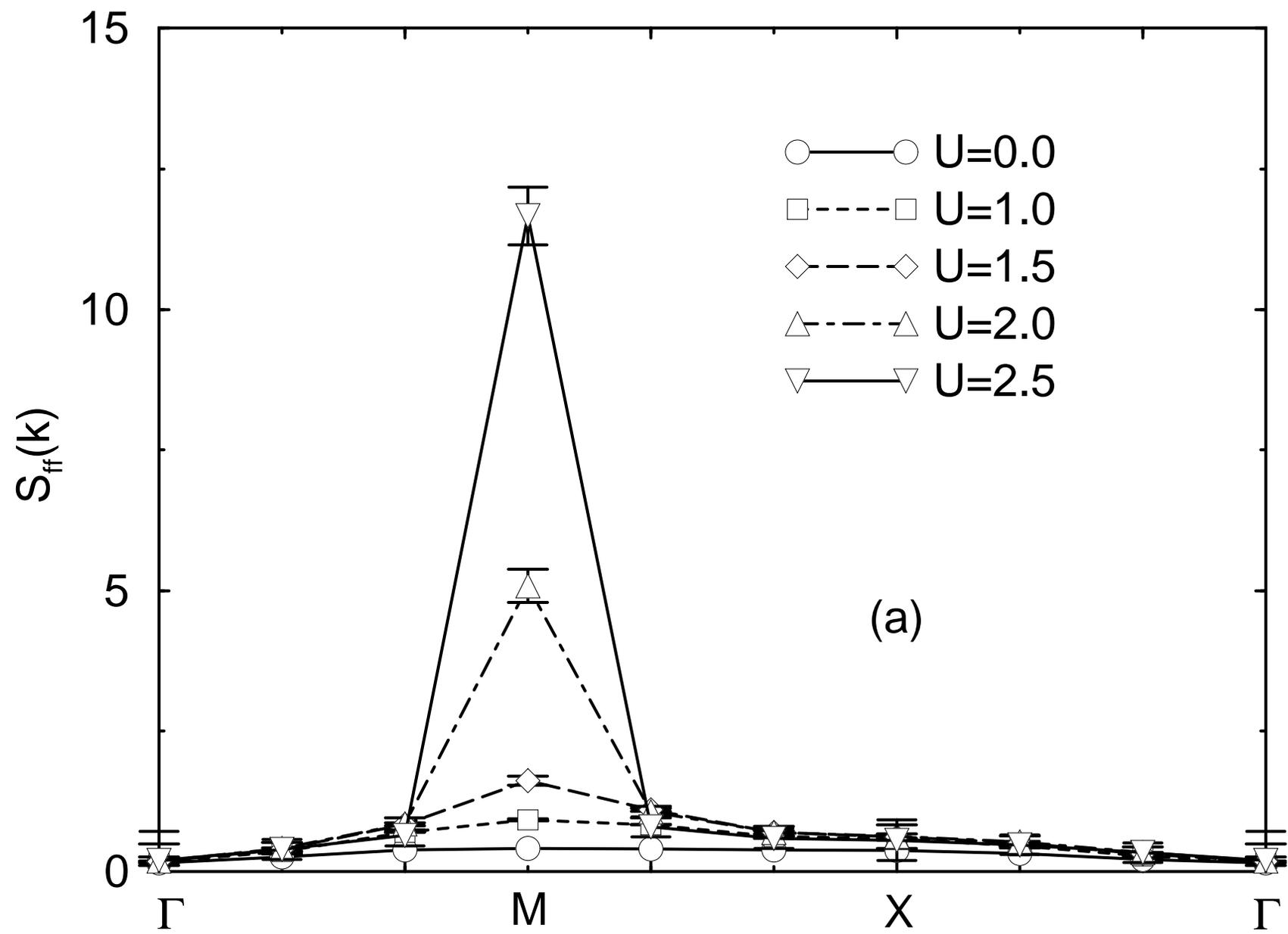
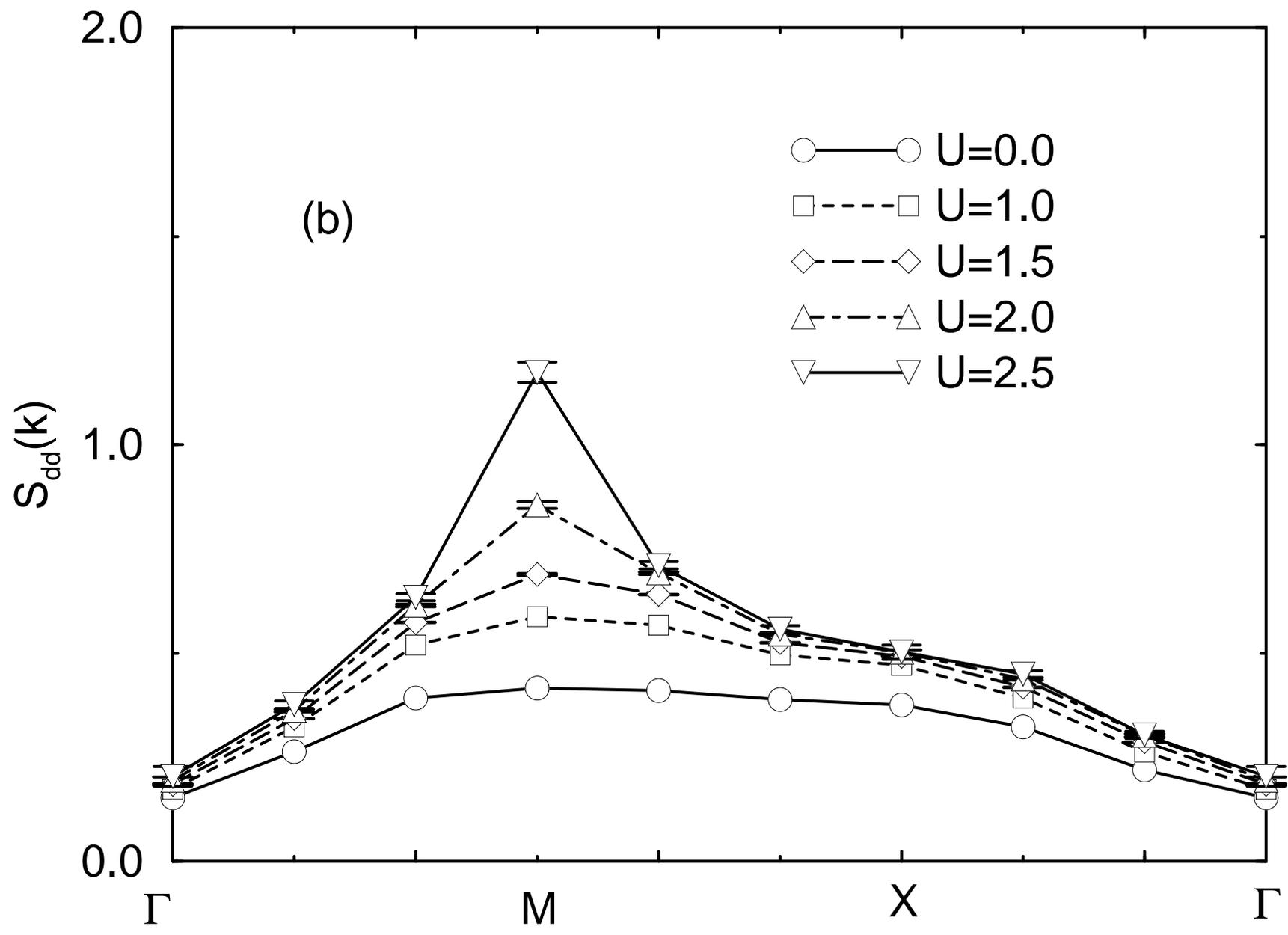
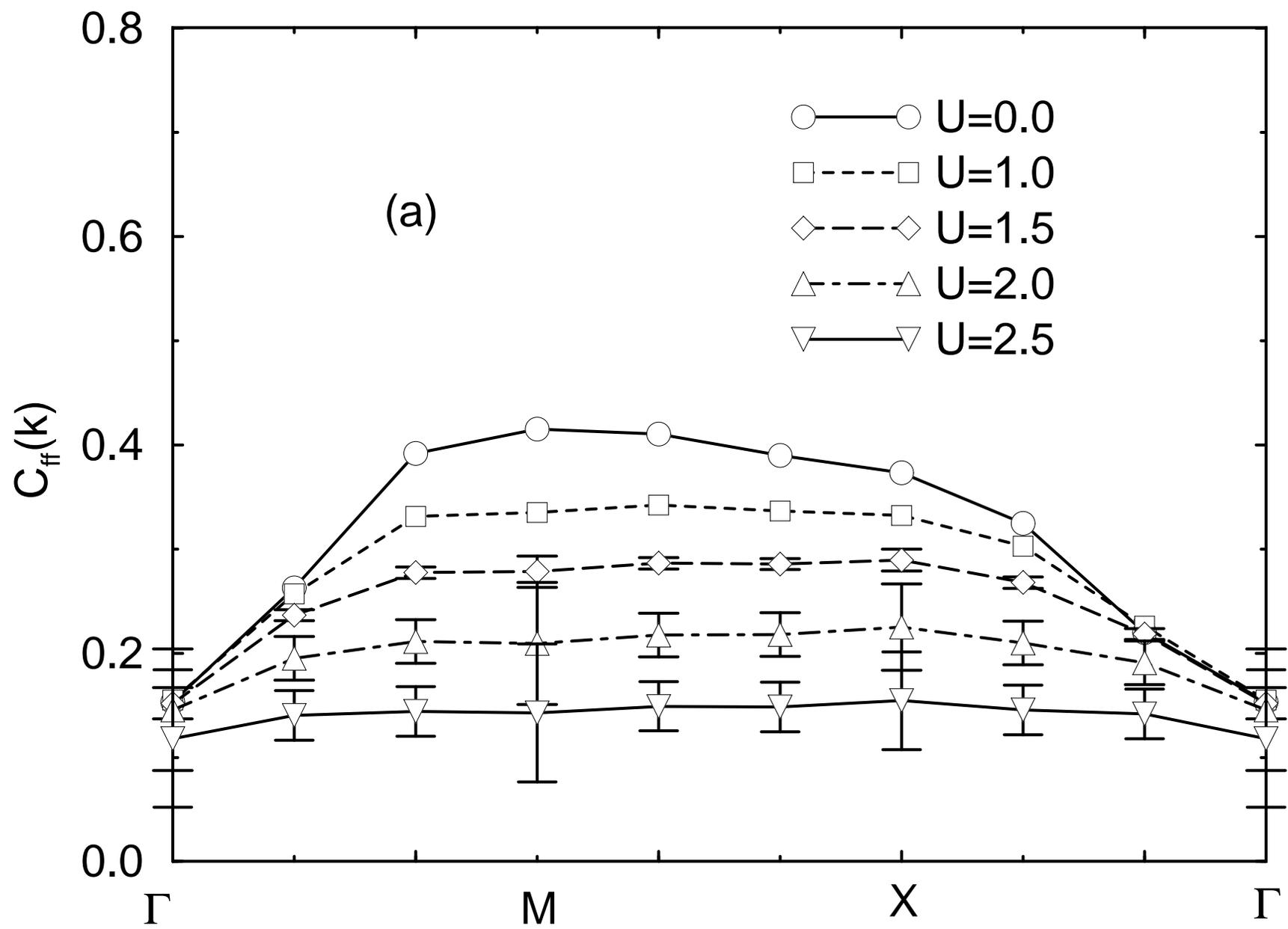
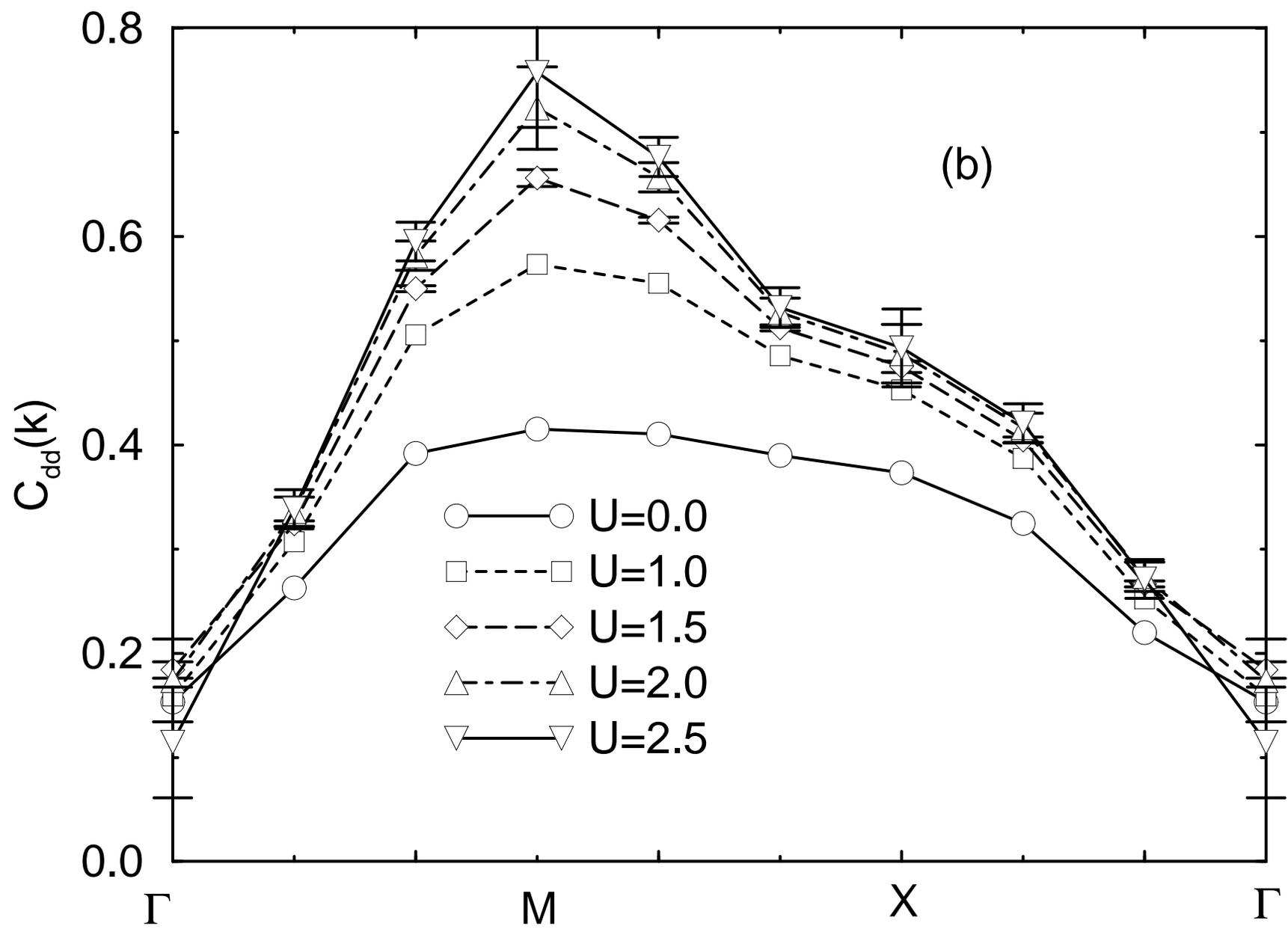
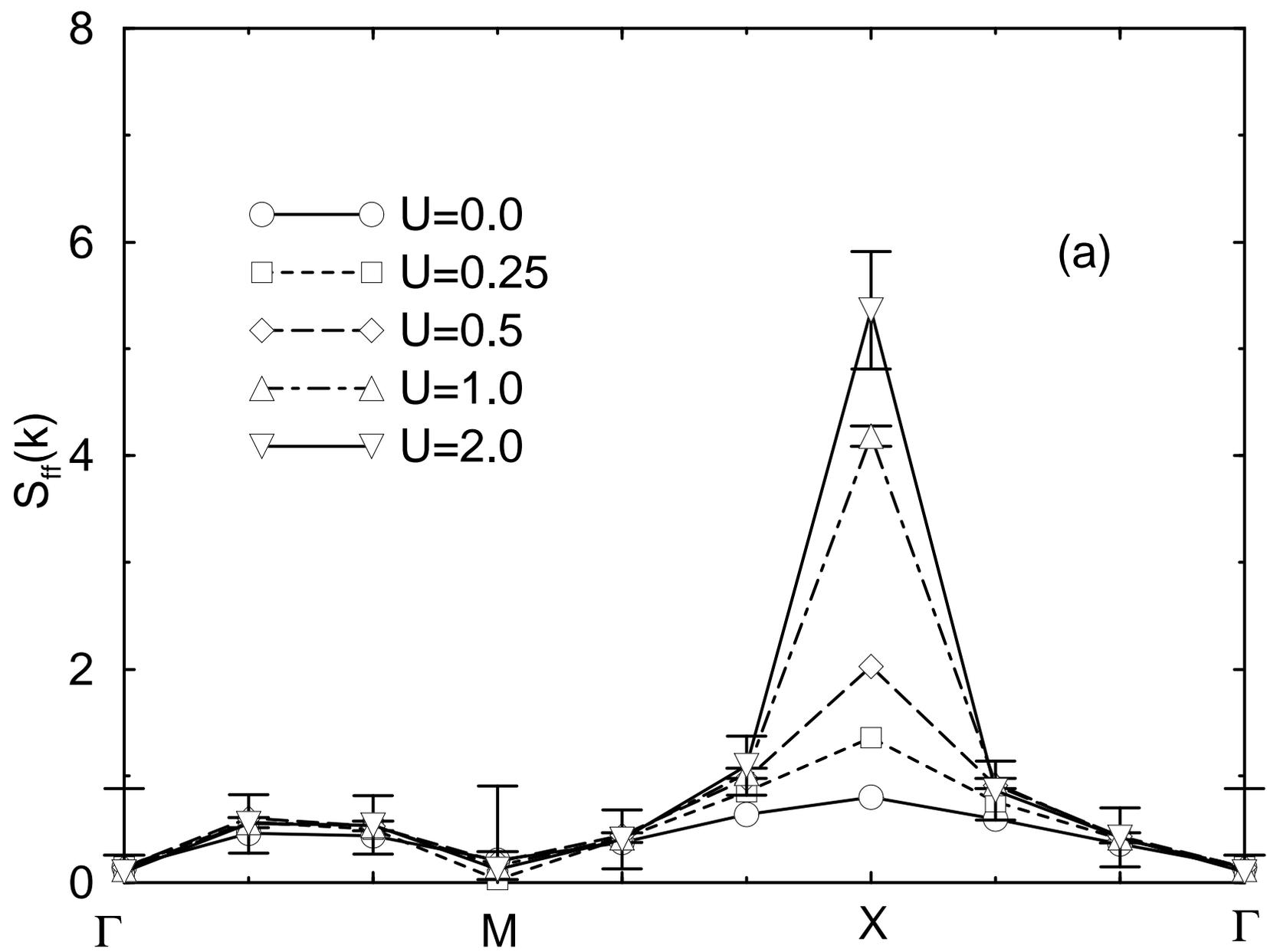
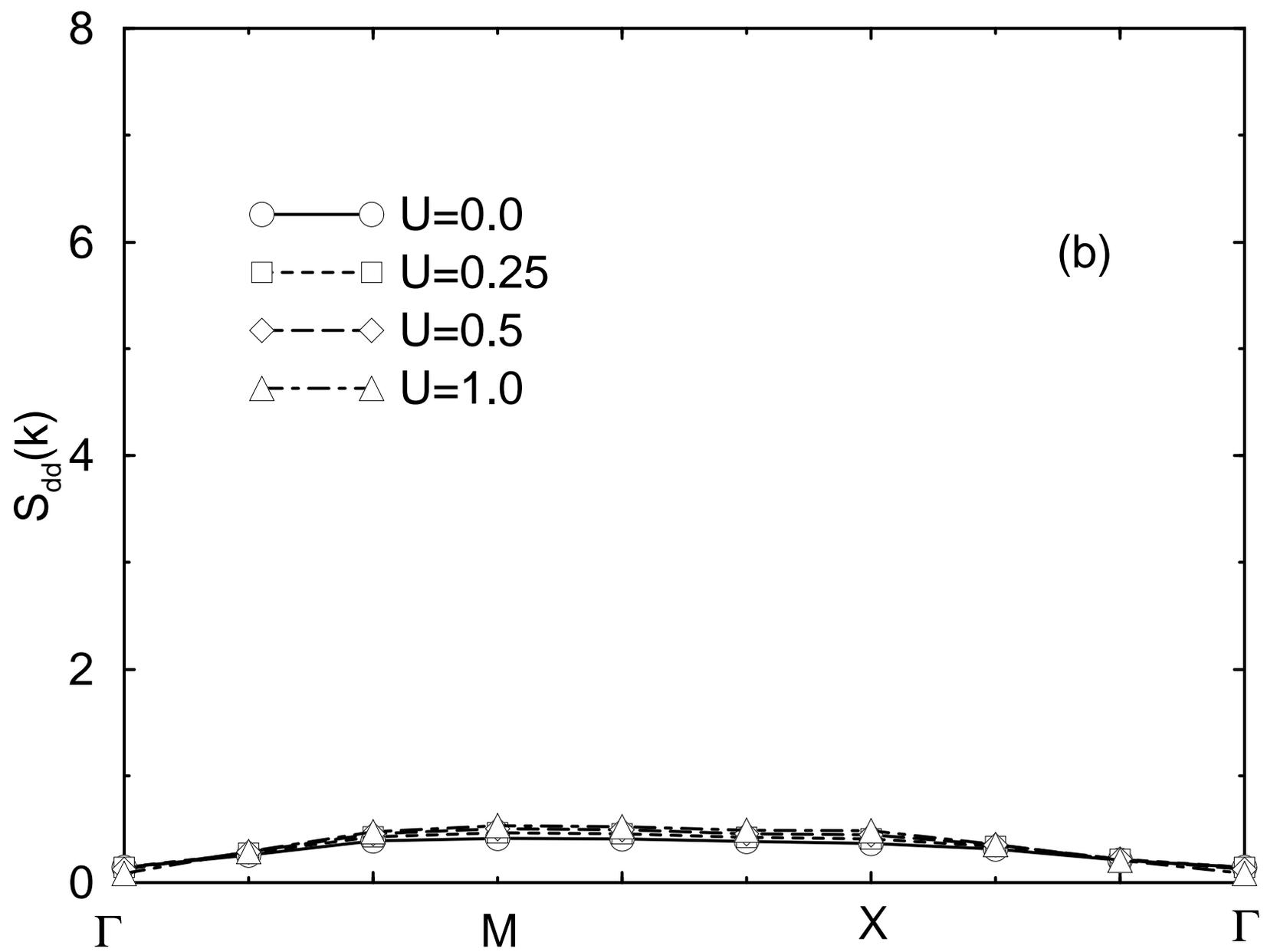
\end{document}